\documentclass{ametsocV6.1}

\usepackage{tabularx,textcomp,graphicx,xcolor}
\usepackage[hidelinks]{hyperref}
\usepackage{booktabs,makecell,array}  
\newcolumntype{C}[1]{>{\centering\arraybackslash}p{#1}}

\newcommand{\eqcontrib}{\textsuperscript{\ensuremath{\dagger}}}
\usepackage[section]{placeins}

\title{Setting the Standard: Recommended Practices for Data Preprocessing in Data-Driven Climate Prediction}
\authors{Jason C. Furtado,\aff{a}\,\eqcontrib\,
\correspondingauthor{Jason C. Furtado, jfurtado@ou.edu} 
Maria J. Molina,\aff{b}\,\eqcontrib\,
Marybeth C. Arcodia,\aff{c}\,\eqcontrib\,
\thanks{\eqcontrib{} These authors contributed equally to this work.}
\thanks{Marybeth C. Arcodia is now at the University of Miami Rosenstiel School of Marine, Atmospheric, and Earth Science Department of Atmospheric Sciences and the Frost Institute for Data Science and Computing in Miami, FL, USA.}
Weston Anderson,\aff{b} 
Tom Beucler,\aff{d} 
John A. Callahan,\aff{e,f} 
Laura M. Ciasto,\aff{g}
Vittorio A. Gensini,\aff{h}
Michelle L'Heureux,\aff{g}
Kathleen Pegion,\aff{a}
Jhayron S. P\'{e}rez-Carrasquilla,\aff{b}
Maike Sonnewald,\aff{i}
Ken Takahashi,\aff{j}
Baoqiang Xiang,\aff{k,l}
and Brian G. Zimmerman\aff{m}
}

\affiliation{\aff{a}{University of Oklahoma, Norman, OK, USA}\\
\aff{b}{University of Maryland, College Park, MD, USA}\\
\aff{c}{Colorado State University, Fort Collins, CO, USA}\\
\aff{d}{University of Lausanne, Lausanne, Switzerland}\\
\aff{e}{Ocean Associates, Inc., Arlington, VA, USA}\\
\aff{f}{National Ocean Service, NOAA, Silver Spring, MD, USA}\\
\aff{g}{NOAA/NWS/NCEP/Climate Prediction Center, College Park, MD, USA}\\
\aff{h}{Northern Illinois University, Dekalb, IL, USA}\\
\aff{i}{University of California, Davis, Davis, CA USA}\\
\aff{j}{Instituto Geofísico del Per\'{u}, Lima, Per\'{u}}\\
\aff{k}{NOAA/GFDL, Princeton, NJ, USA}\\
\aff{l}{Cooperative Programs for the Advancement of Earth System Science, UCAR, Boulder, CO, USA}
\aff{m}{Macquarie, Houston, TX, USA}
}

\abstract{Artificial intelligence (AI) -- and specifically machine learning (ML) --- applications for climate prediction across timescales are proliferating quickly. The emergence of these methods prompts a revisit to the impact of data preprocessing, a topic familiar to the climate community, as more traditional statistical models work with relatively small sample sizes. Indeed, the skill and confidence in the forecasts produced by data-driven models are directly influenced by the quality of the datasets and how they are treated during model development, thus yielding the colloquialism, ``garbage in, garbage out.” As such, this article establishes protocols for the proper preprocessing of input data for AI/ML models designed for climate prediction  (i.e., subseasonal to decadal and longer). The three aims are to: (1) educate researchers, developers, and end users on the effects that preprocessing has on climate predictions; (2) provide recommended practices for data preprocessing for such applications; and (3) empower end users to decipher whether the models they are using are properly designed for their objectives. Specific topics covered in this article include the creation of (standardized) anomalies, dealing with non-stationarity and the spatiotemporally correlated nature of climate data, and handling of extreme values and variables with potentially complex distributions. Case studies will illustrate how using different preprocessing techniques can produce different predictions from the same model, which can create confusion and decrease confidence in the overall process. Ultimately, implementing the recommended practices set forth in this article will enhance the robustness and transparency of AI/ML in climate prediction studies. } 

\begin{document}

\maketitle

%
%
%
\statement
With the rapid expansion of artificial intelligence (AI) in atmospheric science, the need for high quality, properly prepared data for input into AI/ML models is important. In this article, we offer several recommended steps to properly preprocess input data for AI models used for climate predictions (i.e., timescale of a few weeks to many years). Among other topics, we discuss appropriate ways to calculate departures (or anomalies) from data that vary in time and space, how to handle large trends, and what to do with extreme values. We then conduct two case studies to illustrate how using different techniques for preprocessing can produce different predictions from the same model. Ultimately, following these recommendations will help make such studies more transparent, reproducible, and trustworthy. 
%
\capsule
Key recommendations for data preprocessing and problem design in artificial intelligence applications for climate prediction are detailed. Following these recommendations will help make such studies more transparent, reproducible, and trustworthy.
%

%
\section{Introduction}\label{sec:Intro}
The integration of artificial intelligence and machine learning (AI/ML) in weather and climate science is rapidly revolutionizing predictions and our understanding of Earth climate system. Such techniques offer increased capabilities in handling large datasets, identifying complex patterns, and making accurate predictions. Recent attention has been heavily centered on model choice \citep[i.e., selecting which type of ML model is most appropriate; ][]{dueben2018,deburgh-day2023,molina2023review} and explainability of the predictions \citep[e.g.][]{mamalakis2022,mamalakis2023,yik2023,camps-valls2025}. However, the effectiveness of data-driven models strongly depends on data quality. This consideration is paramount and led to the popular adage ``garbage in, garbage out,'' credited to computer programmers in the late 1950s \citep{lidwell2003}. Simply put, if flawed or poor-quality data are fed into a model, the resulting predictions will likely also be flawed. 

Data quality can be evaluated in multiple ways. One method is based on sample size and the fidelity of the data. These considerations are important for AI/ML applications and have been addressed in several works across disciplines \citep[e.g.,][]{dueben2022,deburgh-day2023,zantvoort2024,xie2025}. For the atmospheric sciences, so-called ``benchmark platforms” that provide datasets ready for AI/ML applications have been developed -- e.g., WeatherBench \citep{rasp2020,rasp2024}, AQ-Bench \citep{betancourt2021}, ClimateBench \citep{watson-parris2022}, ClimSim \citep{yu2024}, and ChaosBench \citep{nathaniel2024}. However, even if one has high quality data with sufficient samples, a data-driven model may still produce poor results, as ``garbage” can result from other erroneous assumptions or treatments of the input data: e.g., wrong assumptions around data biases, incorrect assessment of the ``true” distribution, incorrect classification labels, and inconsistent thresholds and definitions of phenomena. 

Climate data present unique challenges for use in data-driven models. Most climate datasets are inherently spatiotemporal, sparse, and possess spatial and temporal autocorrelations. The data are often nonstationary, especially in recent decades, due to anthropogenic climate change. This effect changes the core statistics of the variables of interest (e.g., temperature, wind, cloud cover, geopotential height, sea level rise). Studies handle nonstationarity in different ways. For example, when removing trends from a time series, some studies may simply remove a linear or second-order polynomial trend \citep[e.g.][]{Long2025}, while others may use more complex techniques, such as empirical mode decomposition \citep[e.g.,][]{huang1998}, to detect changing trends over time. Along with changing background statistics, climate variables also have varying distributions, many of which are non-normal (e.g., gamma, bimodal, log-normal, and skew-normal), and exhibit non-linear interactions among themselves. As such, many traditional methods in statistics and AI/ML cannot simply be used ``out of the box” when working with climate variables. Additionally, climate data are collected from diverse sources, including satellite observations, weather stations, and climate models, and can be noisy, incomplete, and heterogeneous. The rapid development of AI/ML applications in climate prediction -- i.e., forecasting the state of the climate system (in probabilities) on timescales ranging from several weeks to a couple of decades in the future -- necessitates a guiding set of recommended preprocessing steps for climate data to secure some degree of harmony between studies and applications. As such, understanding the rationale for \emph{why} certain data preprocessing steps are taken can improve trust in these methods for end users by demystifying the process and providing ways to evaluate and critique the models and their results. 

The aim of this paper is to present recommended practices on proper data preprocessing steps aimed for climate prediction studies using AI/ML. This paper has three overarching goals:

\begin{enumerate}
     \item Educate researchers and end users alike on different effects that dataset preprocessing can have on climate prediction across timescales (i.e., subseasonal to decadal and longer);
     
     \item Provide researchers with recommended practices for dataset preparation in such studies; and
     
     \item Empower end users to determine whether the models they are using are properly designed for their objectives, which can enhance the trustworthiness of AI/ML.
\end{enumerate}

\section{Recommendations for Initial Steps in Data Preparation and Minimizing Data Leakage} \label{sec:initial}
The initial step in climate prediction is to clearly identify what the researcher wants to predict, over which timescale(s) to make this prediction, and what AI/ML methods are most appropriate for addressing the prediction problem. This step is essential and should be carefully considered –- the reader is referred to other works for discussion of problem setup and different AI/ML methods used in prediction studies \citep[e.g.,][]{molina2023review,yang2024,camps-valls2025}. Upon making these decisions, the next step is to select a set of potential input features (i.e., predictors) and outputs used to generate a prediction or classification with that AI/ML model during training or testing. Most prediction problems will use a supervised learning framework requiring inputs and outputs; unsupervised learning applications only need inputs. Inputs and outputs can take the form of a numerical value (e.g., the Ni\~{n}o 3.4 index), numerical data fields (e.g., sea surface temperature), categories (e.g., an El Ni\~{n}o or La Ni\~{n}a event), or a probability. 
More about the terminology above can be found in \citet{chase2022}.

We recommend a period of ``data exploration” first to identify key first-order statistics of the input features and locate any missing or erroneous data. As mentioned, climate variables can possess autocorrelation and covariances between them in space and time. Quantifying these relationships allows us to identify the \emph{effective} sample size $\left(N_\mathrm{eff}\right)$, which can often be smaller than the total number of samples \citep[$N$; e.g.,][]{bretherton1999}. Recognizing this concept is important, as it will impact the choice for the number of training samples needed. Sparse data (e.g., weather stations) introduce additional challenges, such as potential sampling biases. Interpolation or imputation can be used to fill undersampled data, but care must be taken with the choice of method (e.g., bilinear vs.\ piecewise constant). Finally, understanding the distributions and trends of the variables will also inform the preprocessing steps needed (see Section \ref{sec:preprocess}).

After initial data exploration, one should establish a representative 
training dataset, which will be used to ``fit” the model’s parameters (e.g., weights or coefficients). In choosing the number of training samples needed, one should consider: (a) model complexity, (b) available computational resources, (c) number of total input features and their ability to represent a range of possible outcomes, and (d) required/or desired accuracy. Having many input features, however, often requires a larger training sample size for the given model. As such, input features can be reduced using filtering methods (i.e., eliminating features with little-to-no statistical relationship with the target) or feature extraction (i.e., dimensionality reduction using, for example, principal component analysis). We advise considering one or more of these methods for feature selection for one’s AI/ML problem.

Thereafter, data ``splits” need to be decided. For this, the datasets are typically divided into three distinct subsets: (1) training (used for fitting the model); (2) validation (used for adjusting configurable components of the model, known as hyperparameters), and (3) testing (used for final model evaluation). Commonly used ratios for training, validation, and testing sets include 60:20:20 or 80:10:10. Since the validation dataset is used during model development, it should not be considered in assessing test results. Doing so contributes to data leakage (discussed below), potential data misuse, and even unethical use of AI/ML. We further recommend that results for the training, validation, and testing datasets be openly reported to assess, for example, the generalizability properties of the model (e.g., how well the model will perform with the same predictors at a future time). We encourage the use of cross-validation (CV) methods to mitigate learned bias as well as using the full training and validation sets for more robust hyperparameter tuning \citep[e.g.,][]{sweet2023}. $k$-fold cross-validation is the most commonly used cross-validation method in practice; Table \ref{tab:cv} shows other CV methods, along with their limitations.

\begin{figure}
    \centering
    \includegraphics[width=0.9\linewidth, clip]{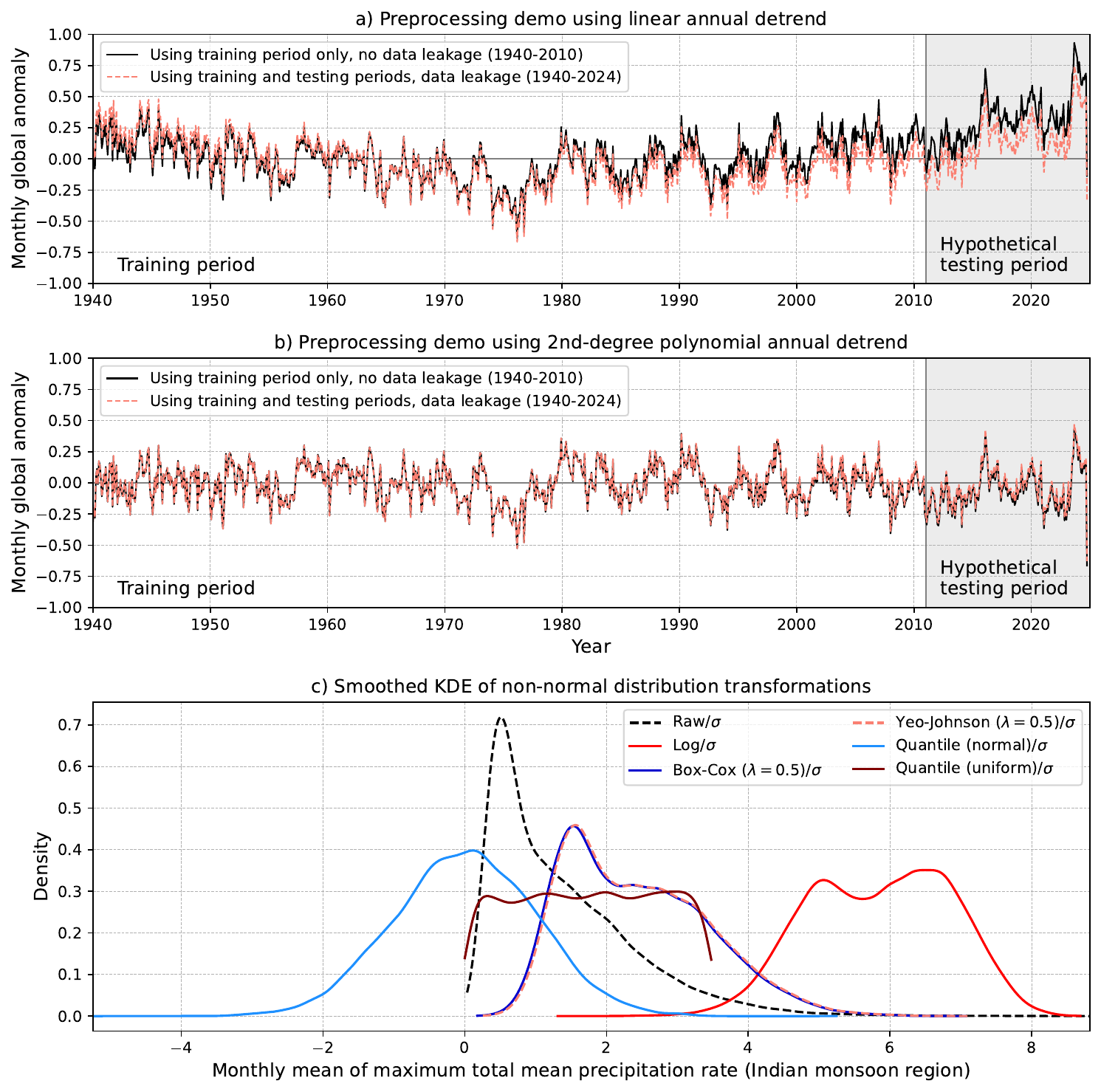}
    \caption{Illustration of various time series preprocessing transformations using ERA5 \citep{hersbach2020}. Preprocessing data leakage for global (latitude weighted) monthly mean of daily maximum temperature anomalies derived from a monthly climatology (1981-2010) is shown in panels (a) and (b), where the black lines represent no data leakage due to detrending using only the training set period, and the pink lines represent leakage due to detrending using the training and (hypothetical) testing periods. Various kernel density estimations of data transformations are shown in panel (c) for a non-normal precipitation variable (black dashed line).} 
    \label{fig:detrend}
\end{figure}

As mentioned, improper splitting choices of the data can inadvertently lead to data leakage, resulting in artificially inflated performance metrics and poor generalization of the results. Leakage of training or testing data occurs when input features contain or are derived from information that can reveal the target variable and would not be (feasibly) available during real-time predictions (i.e., taking information from the future, which one would not have access to at the time of prediction). Data leakage is especially problematic in climate prediction, where variables often possess strong temporal autocorrelation (i.e., ``memory”), quasi-periodicity, and/or long-term trends. Figure \ref{fig:detrend}a

\begin{table}[h!]
    \caption{Cross-validation (CV) methods relevant to weather and climate applications, including example use cases, descriptions, and associated limitations. Types of CV can be combined when the use case has several data properties to consider. In the table, i.i.d. stands for ``independent and identically distributed."}
    \centering
    \fontsize{10pt}{10pt}\selectfont
    \begin{tabular}{>{\centering\arraybackslash}p{0.2\linewidth}>{\centering\arraybackslash}p{0.3\linewidth}>{\centering\arraybackslash}p{0.3\linewidth}}
    
    \toprule
    \textbf{Type of CV} & \textbf{Use Case(s) \& Description} & \textbf{Limitation(s)} \\
    \midrule
    Standard $k$-fold & General data (i.i.d.). Equal-sized random splits. & Leakage is possible (e.g., if there is spatial or temporal autocorrelation).\\
    Stratified $k$-fold& Imbalanced data. Retains class proportions in each fold. & Leakage is possible (e.g., if there is spatial or temporal autocorrelation).\\ 
    Time-series & Sequential data with temporal dependencies. Respects temporal order. & Assumes long-term stationarity and no spatial autocorrelation.\\ 
    Spatial $k$-fold & Spatial autocorrelation in data. Ensures spatially disjoint folds (e.g., with clusters or distance) & Training data may become limited due to spatial constraints. Sensitive to disjoint definition.\\
    Spatial block & Irregularly sampled or sparse spatial data. Data are divided into independent blocks. & Sensitive to block size and placement.\\
    Leave-one-out & Suitable for small data. Each sample serves as validation once. & Computationally expensive for large data. Does not address temporal dependencies.\\
    Monte Carlo & Flexible and randomly repeated splits for general data with specified fold ratios. & Does not address spatial or temporal dependencies. Allows overlapping between folds.\\
    \bottomrule
    \end{tabular}
    \label{tab:cv}
\end{table}
\noindent illustrates an example of data leakage during detrending of a global (latitude weighted) time series of the monthly average of daily maximum temperatures. The use of the full-time period for detrending (1940--2024; Fig.\ \ref{fig:detrend}a, red line) versus the training period (1940--2010; Fig.\ \ref{fig:detrend}a, black line) results in much cooler ground-truth (i.e., target) temperatures during the (hypothetical) test period (2011--2024), which can potentially bias the AI/ML model. The choice of a linear (1-degree) or quadratic (2-degree) polynomial for detrending can also have notable effects, where the 2nd-degree polynomial minimizes data leakage effects during the (hypothetical) test period and removes low-frequency artifacts in the training period (Figs.\ \ref{fig:detrend}a and b). Data leakage can also occur 

\noindent during the scaling and normalizing of the input features. Considering temporal autocorrelation in climate data, we recommend block splitting; i.e., the dataset is split into temporally continuous training/validation and testing blocks or periods (e.g., Fig.\ \ref{fig:detrend}a), with a gap between the two blocks adequately long relative to the timescales of interest to increase independence \citep[e.g.,][]{zhu2023}.

To prevent leakage, we recommend splitting data \emph{before} any preprocessing steps. The testing data should be set aside and not used until model development is finalized. For instance, feature selection and calculation of scaling factors or trends should be made with the training/validation data after the split and later applied to the testing data. For data with strong autocorrelation, quasi-periodicity, and/or trends, data splitting by obeying time series dependencies or phenomena properties (e.g., choosing El Ni\~{n}o ``years” based on season instead of calendar year) allows the model to ``learn” about the evolution of the phenomena, thereby reducing the impact of autocorrelation between the training and testing datasets.  If data splits do not consider the properties of the phenomenon, then learning relies more on autocorrelation, which by definition inflates skill metrics from the model. Since subseasonal and longer temporal dependencies can result in an insufficient $N_\mathrm{eff}$ to represent the processes of interest \citep[e.g.,][]{mayer2024}, other techniques should be considered. For example, using data from long simulations of climate models to develop the initial versions of the AI/ML models and overcome the limitations of short observational records is becoming more common \citep[e.g.,][]{ham2019,riveratello2023}. This practice assumes that climate models adequately represent the phenomena of interest and that fine-tuning procedures can correct existing biases. To address spatial autocorrelation, one can stratify geographically distinct regions and build buffer zones between them. If regions are climatologically distinct and such properties are important for model performance, ensuring that data splits contain samples from respective regions while limiting temporal autocorrelation may be appropriate.  

\newpage

\section{Recommendations for Preprocessing Different Types of Input Features}\label{sec:preprocess}

\subsection{Working with numerical-valued data: Anomalies and standardization}\label{subsec:numerical}
Most problems in climate prediction involve predicting the anomaly (i.e., departure from an average) in a given field, allowing for skill evaluation beyond just the varying climatology \citep{hamill2006}. However, computation of anomalies depends on the problem design, and the computation method can yield different interpretations. A baseline period may be fixed (e.g., 1991–-2020) or centered rolling windows (e.g., $\pm15$ years), the latter of which may address nonstationarity. Choosing a baseline period early in the record may result in artificial skill gains during evaluation due to the effects of climate change \citep{wulff2022}. Choosing a base period later in the record, particularly one with a strong trend, will skew early values. Long-term trends should also be removed to avoid overinflating correlations between input features or relationships derived therein. When calculating anomalies and trends, the spatial element of the climate variables should also be considered. These quantities can be computed using global, regional, or grid cell-specific climatology. A global or regional baseline climatology (latitudinally weighted if using gridded data) may be necessary when spatial anomaly patterns must be preserved (e.g., modes of variability). A grid cell-specific climatology accounts for local variations, which may be useful when interested in strong horizontal gradients (e.g., precipitation anomalies). Altogether, we recommend that there should not be a ``one-size-fits-all” approach to anomaly calculation.

A special case exists when using hindcast simulations (i.e., forecasts made with a model for past events). Relatively short hindcast databases (e.g., 10-20 years long) restrict choices for climatologies, potentially leading to averages that use ``future” periods and artificially inflating prediction skill \citep{risbey2021}. Furthermore, hindcast climatologies are special because they are a function of the initialization date and lead time, so as to avoid model drift and bias. Therefore, when working with hindcast simulations, we suggest computing separate lead time-based climatologies and applying smoothing windows to multiyear averages to address gaps in initialization dates \citep[e.g.,][]{pegion2019}. For ensemble-based hindcasts, the above-suggested recommendations for anomaly and trend calculations work well for single-model analyses. When considering multi-model means, anomalies should be computed for each model separately.

Finally, feature scaling, either normalization (e.g., min-max scaling) or standardization (i.e., $z$-scoring), of data inputs for AI/ML applications must be carefully done. Single fields with large magnitudes and/or multiple fields with varying magnitudes can cause instability and prevent convergence, requiring feature scaling. We recommend that users consistently scale all features while also considering model design when making such a choice. Feature scaling is unnecessary when using tree-based models (e.g., random forests) since they use feature thresholds in their structure but is especially important when using algorithms sensitive to variance, data containing outliers, and when using distance-based models (e.g., $k$-means clustering). Since outliers can disproportionately affect feature scaling, unless one is interested in predicting outlier events, we suggest either removing the outliers or winsorizing the distribution -- i.e., outlier values are reassigned to a specified percentile of the data. 

When dealing with variables that have non-Gaussian distributions, other preprocessing transformations can help align input features and should be performed before feature scaling. Fig.\ \ref{fig:detrend}c provides examples of different transformations done on the distribution of monthly-mean total precipitation maxima in the Indian Monsoon region, depending on user needs: (a) the log transformation (Fig.\ \ref{fig:detrend}c, red line), (b) the Box-Cox \citep[Fig.\ \ref{fig:detrend}c, solid blue line;][]{box1964} or Yeo-Johnson \citep[Fig.\ \ref{fig:detrend}c, dashed red line;][]{yeo2000} transformations, and (c) quantile transformations (Fig.\ \ref{fig:detrend}c, light blue and solid red lines). Feature scaling should then be applied after these transformations. 

\subsection{Working with categorical inputs and outputs}
Sometimes prediction problems require predicting a category or class (e.g., El Ni\~{n}o or La Ni\~{n}a). Several methods should be considered when working with categorical data. If the data have ordered ranks (e.g., labeling 2-m temperatures as below average, average, and above average), then integer encoding, where a unique integer value is assigned to the respective category (e.g., 0, 1, or 2), is appropriate. Classes representing unordered data \citep[e.g., European weather regimes;][]{grams2017} should use one-hot encoding, whereby a single value in the vector is set to one (corresponding to the category), and the others are left as zero. For the seven European weather regimes, for example, a day labeled ``Atlantic Ridge” may be encoded [0,0,1,0,0,0,0,0] \citep[the eighth category representing ``no regime";][]{grams2017}. One-hot encoding prevents the AI/ML model from interpreting the categories as having an ordinal relationship. Binary categorical problems (e.g., severe or non-severe thunderstorms) can choose between integer or one-hot encoding. Missing values can be assigned as a separate category/class (e.g., ``missing”) before encoding, or if there are few missing values, they may be deleted.

Some climate prediction problems, particularly when working with extreme events, may have data with very few events and many ``null” events -- e.g., if classifying days with 2-m temperatures exceeding the 99\textsuperscript{th} percentile, there will be 1 event for every 100 days, on average. In this example, the extreme heat days comprise a \emph{minority class} compared to the non-extreme heat days. Hence, the user is faced with what is known as \emph{class imbalance}, meaning there are very few ``hits” for the AI/ML model on which to train, and thus, overall poor performance by the AL/ML model \citep{molina2023review}. Since minority classes can be important in climate prediction studies, one should consider resampling techniques to address the imbalance. For example, the size of the minority class can be increased by randomly duplicating existing samples or using algorithms to create synthetic samples, known as data augmentation. One such data augmentation method is the Synthetic Minority Oversampling Technique \citep[SMOTE;][]{chawla2002}. Care should be taken when oversampling autocorrelated data; techniques for time series imputation such as time-sliced SMOTE can be used \citep{baumgartner2022}. Undersampling of the majority class can also be employed to reduce the imbalance \citep[e.g.,][]{gensini2021,riveratello2023}. Class weights can also be applied, where larger magnitude weights can be assigned to the minority class for an added penalty in its erroneous classification during training. Importantly, the minority class should contain a diverse set of examples from which to learn. Thus, we recommend the input data to have a good $N_\mathrm{eff}$ as opposed to focusing on the total sample size. 

\section{Putting the Recommendations into Practice: Case Studies}\label{sec:casestudies}
\begin{figure}[h]
    \centering
    \includegraphics[width=0.9\linewidth, clip]{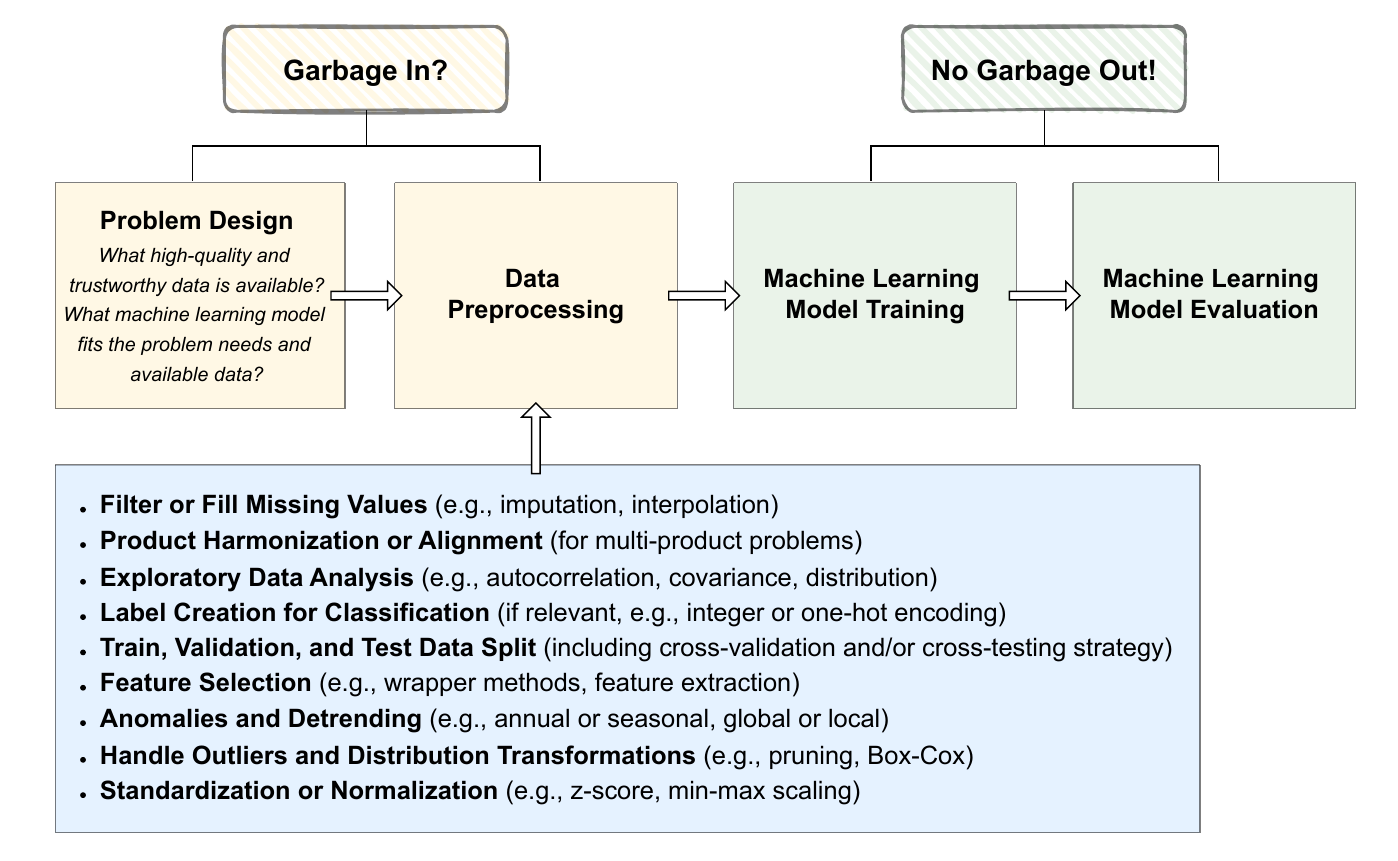}
    \caption{Summary figure illustrating a typical AI/ML workflow, detailing the preprocessing steps for numerical and categorical data, presented in a recommended order. However, preprocessing steps (and their order) may be application-specific.} 
    \label{fig:summary}
\end{figure}

Figure \ref{fig:summary} summarizes the overall workflow for an AI/ML problem in climate prediction, including our recommendations for initial data preparation. We have highlighted the ordering and importance of the preprocessing steps in this flowchart to serve as a template for scientists and practitioners in the field. To further emphasize the importance of these data preprocessing steps, we have designed two small case studies and offer the differences in interpretation and skill that would arise should these steps be followed or not.

\subsection{Case Study \#1: Subseasonal weather regimes}
Weather regimes are large-scale, persistent, and recurrent atmospheric patterns useful for subseasonal predictions due to their relationship with surface weather anomalies \citep[e.g.,][]{molina2023}. These regimes are often defined using 500 hPa geopotential height (Z500) anomalies over a specified domain (e.g., North America) and $k$-means clustering. Here, we demonstrate how seemingly minor differences in preprocessing choices can lead to a different number of preferred weather regimes with different spatial characteristics. To highlight this sensitivity, we will calculate the North American weather regimes in three different ways, detailed below. Daily-mean Z500 comes from ERA5 \citep{hersbach2020} and is regridded to a 1\textdegree \ horizontal grid spacing using the nearest neighbor method. The analysis region is chosen as 20\textdegree N--80\textdegree N, 180\textdegree--30\textdegree W, as in \cite{lee2023}. A daily climatology spanning 1940--2023 is established, utilizing a 60-day centered rolling window method. A 10-day lowpass filter is applied to the anomalies, and they are also detrended by subtracting a latitudinal-weighted third-degree polynomial fit for each day of the year. The data are then standardized using the 60-day centered running mean of the area-averaged Z500 anomaly standard deviation for each day of the year. Subsequent steps are detailed in \citet{perez-carrasquilla2025}.

We conduct three experiments (Table \ref{tab:casestudy1}) for this case study, each with 500 random centroid initializations to ensure robustness of resultant clusters. The ``control” experiment aligns with the definitions and methods used by \cite{lee2023} except for using the 1940–-2023 climatological base period. ``Experiment 1” involves standardizing the detrended anomalies using grid-cell daily-averaged standard deviation instead of using the domain-averaged standard deviation, as done in \cite{lee2023}. ``Experiment 2” involves using the climatological period 1979–-2023. All other preprocessing steps are consistent among the experiments. The optimal number of clusters for each experiment is found using the intercluster correlation, which is the correlation between the centroid coordinates. An intercluster correlation at zero or just below zero is preferred. The optimal number can vary depending on the chosen metric and more than one metric can be used for robustness; however, here we focus on sensitivity to preprocessing choices, not the chosen metric(s). Regime names were assigned subjectively based on similarities in Z500 anomalies, though distance or similarity metrics (e.g., Pearson correlation) can be used for more robust cluster alignment across experiments.

\begin{table}
    \caption{Description of the North American weather regime experiments for Case Study \#1.}
    \centering
    \fontsize{10pt}{10pt}\selectfont
    \begin{tabular}{>{\centering\arraybackslash}p{2.5cm}>{\centering\arraybackslash}p{4cm}>{\centering\arraybackslash}p{2cm}>{\centering\arraybackslash}p{4cm}}
    
    \toprule
    \textbf{Experiment Name} & \textbf{Methodology} & \textbf{Number of Regimes} & \textbf{Regime Names} \\
    \midrule
    Control & \cite{lee2023}, but with 1940–-2023 climatology and 3rd-degree polynomial detrending. & 6 & Alaskan Low (AL), Alaskan Ridge (AR), North Atlantic High (NAH), Pacific Trough (PT), Pacific Ridge (PR), and Greenland High (GH)\\ 
    Experiment 1 & As in control, but with local standardization. & 4 & Pacific Trough (PT), Central US High (CUSH), Pacific Ridge (PR), and Greenland High (GH)\\ 
    Experiment 2 & As in control, but with 1979–-2023 climatology. & 6 & Greenland High (GH), Central US High (CUSH), Alaskan Ridge (AR), Pacific Ridge (PR), Pacific Trough (PT), and North Atlantic High (NAH)\\
    \bottomrule
    \end{tabular}
    \label{tab:casestudy1}
\end{table}

\begin{figure}[h]
    \centering
    \includegraphics[width=0.9\linewidth, clip]{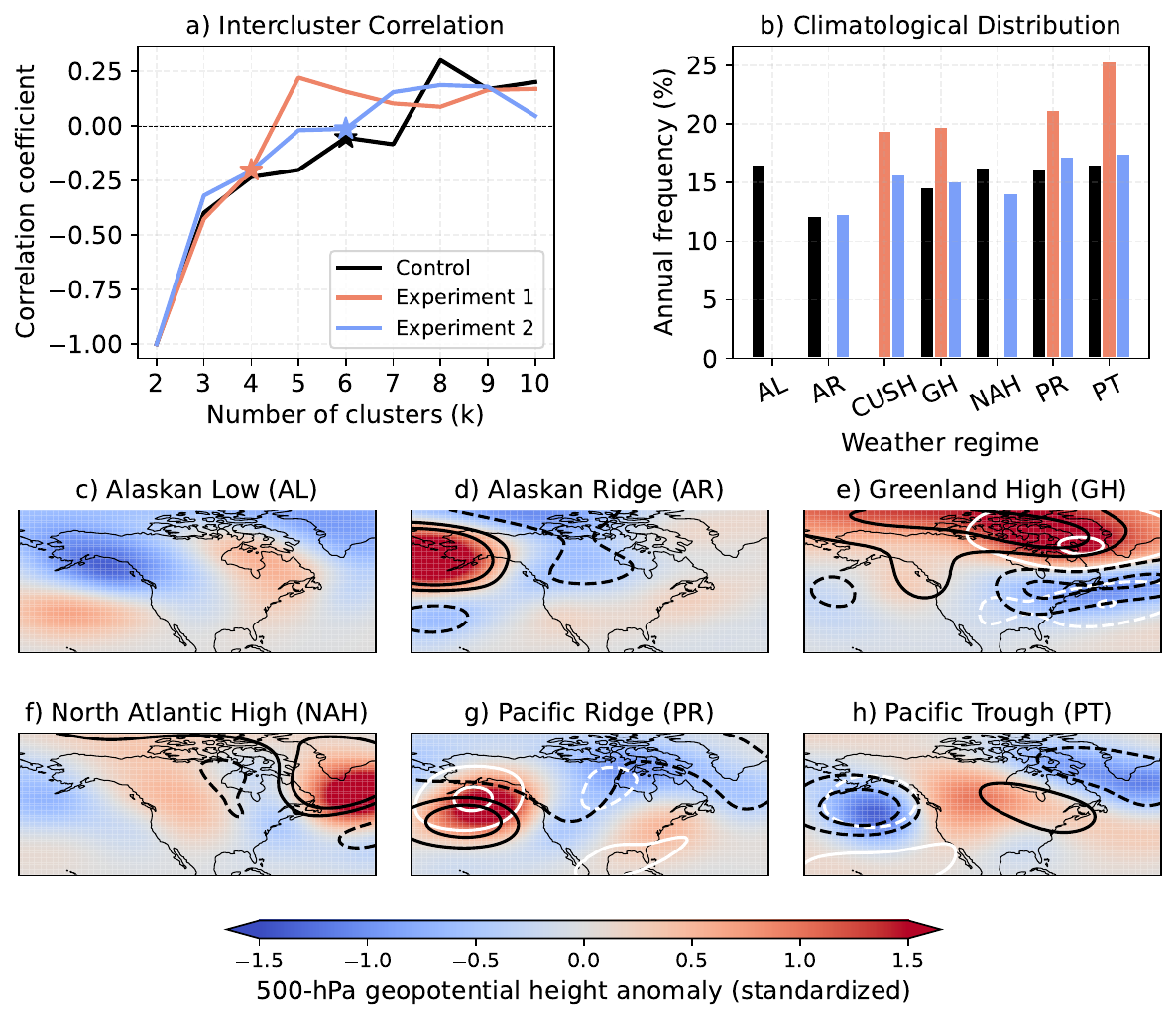}
    \caption{a) Intercluster correlation, with star markers indicating the ``preferred" number of clusters, which keep the correlation between centroids at or just below zero. b) The annual climatological frequency of weather regimes. The control and experiments in panels a) and b) are specified in the legend of panel a). c-h) Standardized Z500 anomalies for the ``control." When the respective regimes were identified in the experiments, contour lines were overlaid. White contour lines represent Experiment 1, while black contour lines denote Experiment 2. Solid lines indicate positive anomalies (+0.5 and +1.0 contour), and dashed lines represent negative anomalies (-0.5 and -1.0).} 
    \label{fig:casestudy1}
\end{figure}

Figure \ref{fig:casestudy1} summarizes the results of the experiments. The control experiment results in the six ``preferred" North American weather regimes due to a relative ``best'' intercluster correlation value at $k=6$ compared to other $k$ values (Fig.\ \ref{fig:casestudy1}a). In contrast, Experiment 1 yields four regimes, whereas Experiment 2 results in six preferred regimes (Fig.\ \ref{fig:casestudy1}a). In Experiment 1, local standardization alters the spatial patterns of the Greenland High, Pacific Ridge, and Pacific Trough regimes (Figs.\ \ref{fig:casestudy1}e, g, h). Local standardization also leads to the emergence of the Central US High regime (not shown), which was absent in the control experiment. In Experiment 2, a shorter climatology eliminates the Alaskan Low regime, suggesting its frequency may have diminished in recent decades (Fig.\ \ref{fig:casestudy1}b). The decision to use grid-cell rather than regional or global standardization stems from the necessity to capture localized signals, potentially useful for subseasonal high-impact weather events. A shorter climatology may be preferred if polynomial detrending is ineffective in removing regional trends. A challenge with unsupervised learning is determining the ``correct" final groupings, but the methodology used in the control experiment is preferred due to the need to capture large-scale patterns (regional standardization) and the intention to investigate trends in weather regimes (longer climatology).

\newpage

\subsection{Case Study \#2: Predicting temperature anomalies in the Southwest US}
Time series prediction is commonly performed in climate science, particularly for understanding and predicting climate patterns. Furthermore, many climate modes and large-scale processes are captured through time series indices, such as the El Ni\~{n}o-Southern Oscillation (ENSO). Forecasting time series evolutions can aid in weather forecasting and prediction of extreme events, such as heat waves. 
To illustrate the impact on prediction skill from common preprocessing missteps, we evaluate a simple neural network regression model to predict the monthly temperature anomaly in the Southwest U.S. Note that the neural network built for this case study is intended to illuminate various preprocessing effects on outcomes and not for actual prediction. For many applications, the observational record is often too short for developing such neural network models, as time series can exhibit strong temporal autocorrelation. Many weather and climate forecasting applications are developed using multi-global climate model data \citep[e.g.,][]{ham2019, riveratello2023}, which increases $N_\mathrm{eff}$ by at least an order of magnitude. However, we have ensured that our model is not overfit to the training data via implementing early stopping (e.g., \texttt{patience}=50, which indicates how many epochs to wait for a model's performance to improve before stopping training) and is sufficient in demonstrating preprocessing differences.  

The case study is set up as follows. Monthly-mean average temperatures are taken from the Berkeley Earth Surface Temperatures dataset \citep{rohde2020} from 1900--2025 and averaged across the Southwest US (29\textdegree N-39\textdegree N; 104\textdegree W-117\textdegree W). We input the current temperature anomaly and three lagged time steps of the timeseries to the neural network (3 layers, 100 nodes in the first layer, 50 nodes each in the second and third layer) to predict the Southwest US temperature anomaly 1 month later. We focus on three specific preprocessing components: 1) the data split, 2) the climatological period, and 3) the detrending period. We construct several variants of the model (i.e., experiments), highlighting the use and misuse of each of these three preprocessing steps (Table \ref{tab:casestudy2}). The \texttt{clean} preprocessing experiment represents the case in which the recommended preprocessing steps highlighted in this article are fully followed. The data split between the training and validation and validation and testing periods is an 18-month gap to reduce data leakage. Additionally, the climatology is computed over a 30-year period from the middle of the training period from January 1941 to December 1970. The linear trend (0.0289\textdegree C decade\textsuperscript{-1}) is also computed over the full training period of January 1900 to December 1979. Skill for all experiments is computed via mean absolute error (MAE) with lower MAE indicating higher skill. We first compute the MAE of the predictions using the test labels as they were originally (incorrectly) computed for each experiment, which results in inflated, ``apparent” skill estimates. We then compute the adjusted MAE using the correctly cleaned test labels, which represents the true prediction error that would be observed in settings with no prior knowledge of the test data, such as real-time forecasting.

The model was trained using the Adam optimizer to minimize mean squared error loss. Additional parameters include a batch size of 64 and a learning rate of 0.00001. Parameters were selected to minimize loss for the \texttt{clean} experiment and the same parameters were used for all experiments. 

\begin{table}[h]
\centering
\small

\setlength{\tabcolsep}{2pt}
\caption{The time periods for each of the data splits and computations with a description for the 5 experiments shown in Figure \ref{fig:casestudy2}.}
\begin{tabularx}{\textwidth}{@{}l >{\centering\arraybackslash}p{0.1\linewidth} >{\centering\arraybackslash}p{0.1\linewidth} >{\centering\arraybackslash}p{0.1\linewidth} >{\centering\arraybackslash}p{0.1\linewidth} >{\centering\arraybackslash}p{0.1\linewidth} >{\centering\arraybackslash}p{0.35\linewidth}@{}}
\toprule
\makecell{\textbf{Experiment}} & \makecell{\textbf{Training}} & \makecell{\textbf{Validation}} & \makecell{\textbf{Testing}} & \makecell{\textbf{Climo}} & \makecell{\textbf{Trend}} & \makecell{\textbf{Description}} \\
\midrule

\makecell{\texttt{clean}} &
\makecell{1900-01-01\\1979-12-01} &
\makecell{1981-07-01\\2001-02-01} &
\makecell{2002-09-01\\2024-12-01} &
\makecell{1941-01-01\\1970-12-01} &
\makecell{1900-01-01\\1979-12-01} &
\makecell{No data leakage; 18-month gap between\\splits}\\

\makecell{\texttt{trend}} &
\makecell{1900-01-01\\1979-12-01} &
\makecell{1981-07-01\\2001-02-01} &
\makecell{2002-09-01\\2024-12-01} &
\makecell{1941-01-01\\1970-12-01} &
\makecell{1900-01-01\\2024-12-01} &
\makecell{Trend computed during\\the entire period}\\

\makecell{\texttt{climo}} &
\makecell{1900-01-01\\1979-12-01} &
\makecell{1981-07-01\\2001-02-01} &
\makecell{2002-09-01\\2024-12-01} &
\makecell{1991-01-01\\2020-12-01} &
\makecell{1900-01-01\\1979-12-01} &
\makecell{Climatology computed\\during test period\\\mbox{}}\\

\makecell{\texttt{split}} &
\makecell{1900-01-01\\1999-12-01} &
\makecell{2000-01-01\\2002-08-01} &
\makecell{2002-09-01\\2024-12-01} &
\makecell{1941-01-01\\1970-12-01} &
\makecell{1900-01-01\\1979-12-01} &
\makecell{Train/val/test split on\\sequential months and\\during ongoing ENSO\\events}\\

\makecell{\texttt{split\_}\\\texttt{trend\_climo}} &
\makecell{1900-01-01\\1979-12-01} &
\makecell{1980-01-01\\2002-08-01} &
\makecell{2002-09-01\\2024-12-01} &
\makecell{1991-01-01\\2020-12-01} &
\makecell{1900-01-01\\2024-12-01} &
\makecell{Trend, climatology com-\\puted during test period;\\split is sequential months}\\
\bottomrule
\end{tabularx}
\label{tab:casestudy2}
\end{table}

\begin{figure}[h]
    \centering
    \includegraphics[width=0.9\linewidth,trim={1.2in 0 0.9in 0},angle=270,scale=0.6,clip]{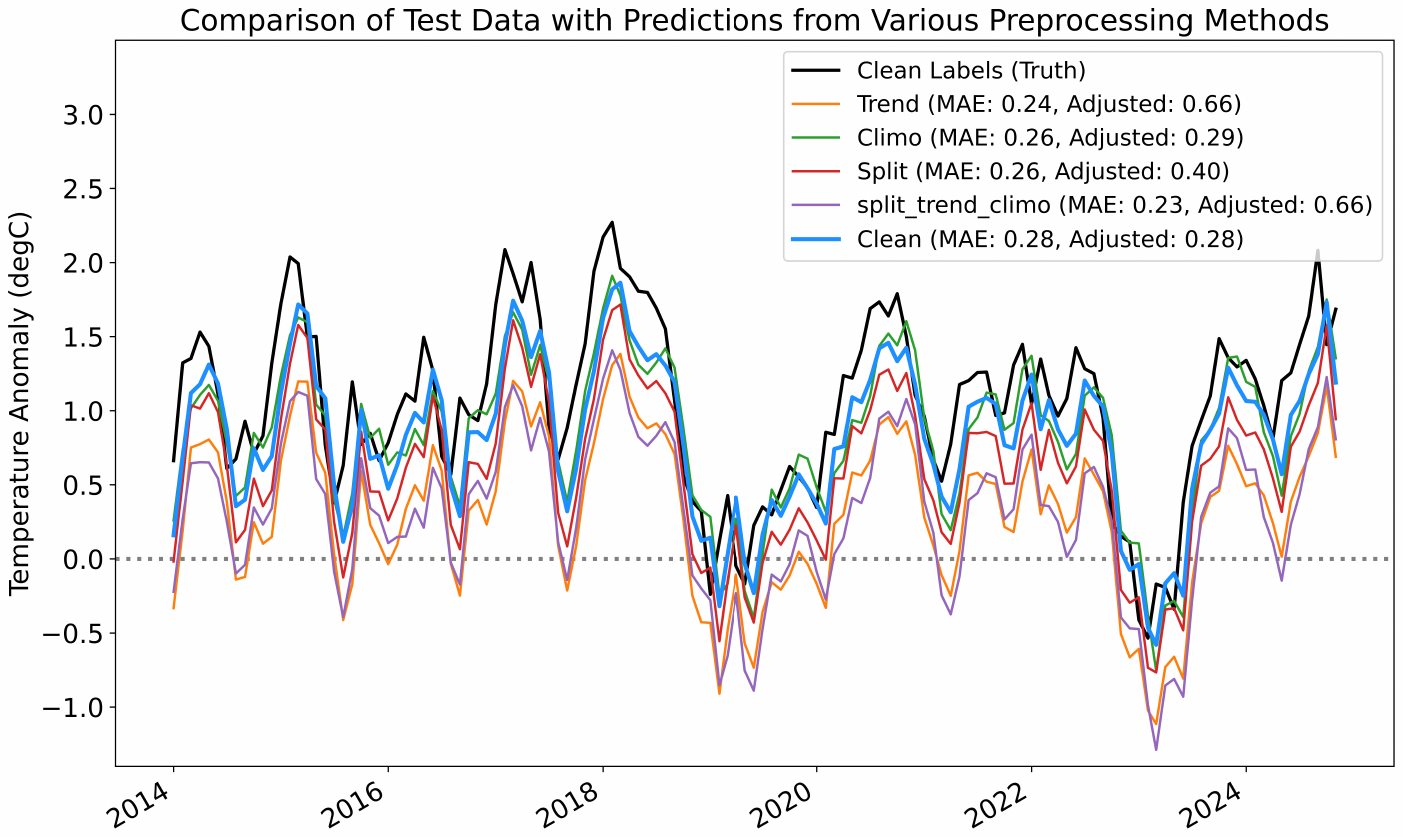}
    \caption{Temperature anomaly (black; the ``truth") and neural network predictions using test datasets preprocessed in different ways: 
    (1) trend computed during the test period (\texttt{trend}; orange), 
    (2) climatology computed during the validation and test periods (\texttt{climo}; green), 
    (3) data splits with potential leakage (\texttt{split}; red), and 
    (4) a combination of the trend, climo, and split preprocessing steps (\texttt{split\_trend\_climo}; purple)
    (5) no data leakage (\texttt{clean}; blue). 
    Corresponding skill scores are shown in parentheses. The MAE (mean absolute error; \textdegree C) is first calculated using the incorrectly computed test labels for each experiment, producing inflated skill estimates. The "adjusted" MAE is calculated using the properly cleaned test labels to obtain an accurate measure of prediction error.
    }
    \label{fig:casestudy2}
\end{figure}

The predictions from the \texttt{clean} experiment and its corresponding skill is shown in blue in Figure \ref{fig:casestudy2}. Comparisons between experiments are summarized below. 
\begin{itemize}

\item For the \texttt{trend} experiment (Fig.\ \ref{fig:casestudy2}, orange line), the linear trend was computed over the full dataset (0.1048\textdegree C decade\textsuperscript{-1}; January 1900 to December 2024). The trend over the full data set is nearly 3 times higher than the trend over the training period. Thus, the error is lower for the trend predictions, due to knowledge of the increased trend during model training. However, this trend from the testing period would not be known in a true testing sense where the testing data are completely unseen. Thus, the model’s performance is inflated, as shown by the adjusted MAE being much higher; the trend has the largest impact on the overall skill in this case study. 
\item For the \texttt{climo} experiment (Fig.\ \ref{fig:casestudy2}, green line), we use climatology calculated from January 1991 to December 2020 (i.e., spanning both the validation and testing periods). We find the error is slightly lower than the \texttt{clean} experiment, but again, this skill is inflated due to knowledge of information from the test data.  
\item For the \texttt{split} experiment (Fig.\ \ref{fig:casestudy2}, red line), we split the training, validation, and testing sets by only a one month separation instead of splitting the data with a sizable gap between the next dataset to avoid data leakage from low frequency variability, The resulting error is lower than the \texttt{clean} experiment, but the skill is again inflated due to data leakage.
\item Finally, for the \texttt{split-trend-climo} experiment (Fig.\ \ref{fig:casestudy2}, purple line), we combine the preprocessing missteps from the three previous experiments. The result is a slightly lower error than those three experiments, resulting in artificially inflated skill when compared with the adjusted MAE.
\end{itemize}

\section{Conclusions}
The use of AI/ML in climate prediction is rapidly expanding, introducing challenges with model design, skill assessment, and ultimately trust. A useful step towards building that trust is transparency in the process, including the initial problem design and data preprocessing. This work presents recommended steps for proper dataset preprocessing for different climate prediction problems. Such steps are recommended across applications and will serve as a way to make conscious choices when framing a prediction problem for climate timescales. The two case studies presented illustrate the importance of our recommended preprocessing steps and show how they can affect interpretation of the predictions. Understanding the importance of these preprocessing steps will likely lessen the frequent critique of the ``black box” nature of AI/ML \citep{mcgovern2019}. Hence, following the recommendations laid out in this article will move the community toward AI/ML applications for climate prediction that are transparent and fairly evaluated. 

While important, these recommended practices are not a replacement for co-production of knowledge in AI/ML. Uncertainties, biases, and other unknowns in climate prediction studies require further work. Moreover, the data preprocessing steps presented here are not a complete substitute for the need to engage with domain experts and stakeholders alike in selecting the appropriate datasets, methods, and verification metrics. Thus, we encourage further collaboration between academics, operational forecasters, and industry scientists to ensure the model predictions are transparent, reproducible, and actionable. This collaboration and transparency includes ensuring any code for the model and the preprocessing steps be openly available. We also advocate for similar recommendations for benchmarking and evaluating AI/ML predictions used in subseasonal-to-seasonal and seasonal-to-decadal timescales. Open sourcing of all recommendations and associated software, from preprocessing to evaluation, would provide the community with an end-to-end roadmap to using AI/ML for a variety of climate prediction problems across scales and applications.

\clearpage
\acknowledgments
The authors would like to thank the US Climate Variability and Predictability Program (CLIVAR) for its support for this paper. M.\ J.\ M.\ was supported by the National Science Foundation (NSF) Grant \#2425735. M.\ C.\ A.\ was funded, in part, by the National Oceanic and Atmospheric Administration (NOAA) Grant \#NA24OARX431C0065-T1-01 and by the US Department of Energy’s (DOE) Office of Biological and Environmental Research (BER) Regional and Global Model Analysis (RGMA) program, as part of the Program for Climate Model Diagnosis and Intercomparison (PMCDI) Project. J.\ S.\ P.\ C.\ was supported by a University of Maryland Grand Challenges Seed Grant. T.\ B.\ received support from from the Horizon Europe project ``Artificial Intelligence for enhanced representation of processes and extremes in Earth System Models (AI4PEX)” (Grant agreement ID: 101137682), funded by the Swiss State Secretariat for Education, Research and Innovation (SERI, Grant No.\ 23.00546). The scientific results and conclusions, as well as any view or opinions expressed herein, are those of the authors and do not necessarily reflect the views of NWS, NOAA, or the Department of Commerce.\bigskip\bigskip

%
%
\datastatement
ERA5 used in Case Study \#1 can be obtained from the US National Science Foundation (NSF) National Center for Atmospheric Research (NCAR) Research Data Archive (\href{https://doi.org/10.5065/D6X34W69}{\textcolor{blue}{https://doi.org/10.5065/D6X34W69}}). Case study \#1 and Figure \ref{fig:detrend} software is publicly available at: \href{https://github.com/jhayron-perez/WRs_Preprocessing_BAMSPaper}{\textcolor{blue}{https://github.com/jhayron-perez/WRs\_Preprocessing\_BAMSPaper}}. Case study \#2 software is publicly available at: \href{https://github.com/mbarcodia/bams-preprocess}{\textcolor{blue}{https://github.com/mbarcodia/bams-preprocess}}.








%



\newpage
\bibliographystyle{ametsocV6}
\bibliography{references}

\end{document}